\documentclass[onecolumn,tightenlines,showpacs,nofootinbib,preprintnumbers,
amsmath,amssymb]{revtex4}

\usepackage[usenames,dvipsnames]{color}
\usepackage{dcolumn}
\usepackage{graphicx}
\usepackage{subfigure}
\usepackage{epstopdf}
\usepackage{bm}
\usepackage{lscape}
\renewcommand{\arraystretch}{0.9}

\usepackage{setspace}

\begin{document}
\begin{spacing}{1.5}

\title{Direct $CP$ violation for $\bar{B}_s^0 \to \phi {\pi^+}{\pi^-}$ in Perturbative QCD}

\author{Sheng-Tao Li$^{1}$\footnote{lst@mails.ccnu.edu.cn},
Gang L\"{u}$^{2}$\footnote{ganglv66@sina.com}
}

\affiliation{
\small $^{1}$Institute of Particle Physics and Key Laboratory of Quark and Lepton Physics~(MOE), \\
\small Central China Normal University, Wuhan, Hubei 430079, China \\
\small  $^{2}$College of Science, Henan University of Technology, Zhengzhou 450001, China}

\begin{abstract}
In perturbative QCD approach, we study the direct $CP$ violation in the decay of $\bar{B}_s^0\to\rho(\omega )\phi\to {\pi^+}{\pi^-}\phi$ via isospin symmetry breaking.
An interesting mechanism involving the charge symmetry violating between $\rho$ and $\omega$ is applied to enlarge the $CP$ violating asymmetry. We find that the $CP$ violation can be
enhanced by the $\rho-\omega$ mixing mechanism when the invariant masses of the $\pi^+\pi^-$ pairs are in the vicinity of the $\omega$ resonance. For the decay process of $\bar{B}^0_{s}\to\rho(\omega )\phi\to{\pi^+}{\pi^-}\phi$, the maximum $CP$ violation can reach $5.98\%$. The possibility of detecting the $CP$ violation is also presented.
\end{abstract}


\maketitle

\section{\label{intro}Introduction}

$CP$ violation has obtained extensive attention even since its first discovery in the  Neutral kaon systems \cite{Christenson:1964fg}. Within the standard model (SM), $CP$ violation originates from a non-zero weak phase angle from the complex Cabibbo-Kobayashi-Maskawa (CKM) matrix, which describes the mixing of weak interaction and mass eigenstates of the quarks \cite{Cabibbo:1963yz,Kobayashi:1973fv}. Although the source of $CP$ violation has not been well understood up to now, physicists are striving to increase their knowledge of the mechanism for the $CP$ violation. Many theoretical studies \cite{Carter:1980hr,Carter:1980tk,Bigi:1981qs} (within and beyond the SM) and experimental investigations have been conducted since 1964. According to theoretical predictions, large $CP$ violation may be expected in $B$ meson decay process due to the large mass of b quark. In recent years, the LHCb Collaboration observed the large $CP$ violation in the three-body decay channels of $B^{\pm}\rightarrow \pi^{\pm}\pi^{+}\pi^{-}$ and $B^{\pm}\rightarrow K^{\pm}\pi^{+}\pi^{-}$ \cite{Aaij:2013bla,Aaij:2014iva,Aaij:2013sfa}. Hence, the non-leptonic $B$ meson three-body decays channels have been focused more attention in searching for $CP$ violation.

Direct $CP$ violation in $B$ meson decays process occurs through the interference of at least two amplitudes with different weak phase $\phi$ and strong phase $\delta$. The weak phase difference $\phi$ is determined by the CKM matrix elements, while the strong phase can be produced by the hadronic matrix elements and interference between the intermediate states. However, one can know that the strong phase $\delta$ is not still well determined from the theoretical approach. The non-leptonic weak decay amplitudes  of $B$ meson involve the hadronic matrix elements of $\left\langle {{M_1}{M_2}} \right|{O_i}\left| B \right\rangle $, which can be calculated from the different factorization methods. However, the different methods may present different strong phases so as to affect the value of the $CP$ violation. Currently, there are three popular theoretical approaches to study the dynamics of the two-body hadronic decays, which are the naive factorization approach \cite{Fakirov:1977ta,Cabibbo:1977zv,Wirbel:1985ji,Bauer:1986bm}, the QCD factorization (QCDF) \cite{Beneke:1999br,Beneke:2000ry,Sachrajda:2001uv,Beneke:2001ev,Beneke:2003zv}, perturbative QCD (pQCD) \cite{Lu:2000em,Keum:2000ph,Keum:2000wi}, and soft-collinear effective theory (SCET) \cite{Bauer:2000yr,Bauer:2001cu,Bauer:2001yt}. Based on the power expansion in $1/m_b$ ($m_b$ is the b-quark mass), all of the theories of factorization are shown to deal with the hadronic matrix elements in the leading power of $1/m_b$. However, these methods pertain to whether one takes into account the collinear degrees of freedom or the transverse momenta. Meanwhile, in order to have a large signal of $CP$ violation, we need appeal to some phenomenological mechanism to obtain a large strong phase $\delta$. $\rho-\omega$ mixing has been used for this purpose in the past few years  and focuses on the naive factorization and QCD factorization approaches \cite{Enomoto:1996cv,Guo:1998eg,Guo:2000uc,Leitner:2002xh,Lu:2010vb,Lu:2011zzf}. Recently, L$\rm{\ddot{u}}$ et al. attempted to generalize the pQCD approach to the three-body non-leptonic decay via $\rho-\omega$ mixing in ${B^{0, \pm }} \to {\pi ^{0, \pm }}{\pi ^ + }{\pi ^ - }$ and $B_{c}\rightarrow D_{(s)}^+\pi^+\pi^-$ decays \cite{Lu:2013jma,Lu:2016lgc}. In this paper,
 we will focus on the $CP$ violation of the decay process $\bar{B}^0_{s}\to\rho(\omega )\phi\to{\pi^+}{\pi^-}\phi$ via $\rho-\omega$ mixing in the pQCD approach.

Isospin symmetry breaking plays a significant role in the $\rho-\omega$ mixing mechanism. The mixing between the $u$ and $d$ flavors leads to  the breaking of isospin symmetry for the $\rho-\omega$ system \cite{Fritzsch:2000pg,Fritzsch:2001aj}. In Refs. \cite{Gardner:1997ie,OConnell:1995nse}, the authors studied  the $\rho-\omega$ mixing and the pion form factor in the time-like region, where $\rho-\omega$ mixing was used to obtain the (effective)  mixing  matrix element $\widetilde{\Pi}_{\rho\omega}(s)$, which consists of two part contributions: one from the direct coupling of $\omega  \to 2\pi $ and the other from $\omega  \to \rho  \to 2\pi$ mixing \cite{Bernicha:1994re,Maltman:1996kj,OConnell:1997ggd}. The magnitude has been determined by the pion form factor
through the data for the cross section of ${e^ + }{e^ - } \to {\pi ^ + }{\pi ^ - }$ in the $\rho$ and $\omega$ resonance region \cite{OConnell:1995nse,OConnell:1996amv,OConnell:1997ggd,Wolfe:2009ts,Wolfe:2010gf}. Recently, isospin symmetry breaking was discussed by incorporating the vector meson dominance (VMD) model in the weak decay process of the  meson \cite{Gardner:1997yx,Guo:1999ip,Guo:2000uc,Lu:2016lgc,Lu:2014uja}. However, one can find that $\rho-\omega$ mixing produces the large $CP$ violation from the effect of isospin symmetry breaking in the three and four bodies decay process. 
Hence, in this paper, we shall follow the method of Refs. \cite{Gardner:1997yx,Guo:1999ip,Guo:2000uc,Lu:2016lgc,Lu:2014uja} to investigate the decay process of $\bar{B}^0_{s}\to\rho(\omega )\phi\to{\pi^+}{\pi^-}\phi$ via isospin symmetry breaking.

The remainder of this paper is organized as follows. In Sec. \ref{sec:hamckm} we will briefly introduce the pQCD framework and present the form of the effective Hamiltonian and wave functions.
In Sec. \ref{sec:cpv1} we give the calculating formalism and details of the $CP$ violation from $\rho-\omega$ mixing
in the decay process $\bar{B}_s^0\to\rho(\omega )\phi\to {\pi^+}{\pi^-}\phi$. In Sec. \ref{int} we show the input parameters.
We present the numerical results in Sec. \ref{num}.
Summary and discussion are included in
Sec. \ref{sec:conclusion}. The related function defined in the text are given
in Appendix.

\section{\label{sec:hamckm}The FrameWork}

For the decay process of $\bar{B}_{s} \to {M_2}{M_3}$, integrated over the longitudinal and the
transverse momenta, the emitted or annihilated particle $M_2$ can
be factored out. The rest of the amplitude can be expressed as the convolution
of the wave functions $\phi_{B_s}$, $\phi_{M_3}$ and the hard
scattering kernel $T_H$.
The pQCD factorization theorem has been developed for
non-leptonic heavy meson decays, based on the formalism
of Lepage, Brodsky, Botts and Sterman \cite{Chang:1996dw,Yeh:1997rq,Lepage:1980fj,Botts:1989kf}.
The basic idea of the pQCD approach is that it takes into account the
transverse momentum of the valence quarks in the hadrons which results in the Sudakov
factor in the decay amplitude. Then, it is conceptually written as the following:
\begin{eqnarray}
 \mbox{Amplitude}
\sim \int\!\! d^4k_1 d^4k_2 d^4k_3\ \mathrm{Tr} \bigl[ C(t)
\phi_{B_s}(k_1) \phi_{M2}(k_2) \phi_{M3}(k_3) T_H(k_1,k_2,k_3, t)
\bigr],  \label{eq:convolution1}
\end{eqnarray}
where $k_i(i=1,2,3)$ are momenta of light quarks in the mesons.
 $\mathrm{Tr}$ denotes the trace over Dirac and color indices.
$C(t)$ is Wilson coefficient which comes from the radiative
corrections at short distance. $\phi_{M}(m=2,3)$ is wave function which
describes non-perturbative contribution during the hadronization of mesons, which should
be universal and channel independent. The hard part $T_H$ is rather process-dependent.

With the operator product expansion, the effective weak Hamiltonian in bottom hadron decays is \cite{Buchalla:1995vs}
\begin{eqnarray}
 {\cal H}_{eff} &=& \frac{G_{F}}{\sqrt{2}}
     \Bigg\{ V_{ub} V_{us}^{\ast} \Big[
     C_{1}({\mu}) Q^{u}_{1}({\mu})
  +  C_{2}({\mu}) Q^{u}_{2}({\mu})\Big]
  -V_{tb} V_{ts}^{\ast} \Big[{\sum\limits_{i=3}^{10}} C_{i}({\mu}) Q_{i}({\mu})
  \Big ] \Bigg\} + \mbox{H.c.} ,
 \label{2a}
 \vspace{2mm}
 \end{eqnarray}
where $G_F$ is the Fermi constant, $C_i(\mu)$ (i=1,...,10) are the Wilson coefficients,
$V_{q_1q_2}$ ($q_1$ and $q_2$ represent quarks) is the CKM matrix element. The
operators $O_i$ have the following forms:
\begin{eqnarray}
O^{u}_1&=& \bar s_\alpha \gamma_\mu(1-\gamma_5)u_\beta\bar
u_\beta\gamma^\mu(1-\gamma_5)b_\alpha,\nonumber\\
O^{u}_2&=& \bar s \gamma_\mu(1-\gamma_5)u\bar
u\gamma^\mu(1-\gamma_5)b,\nonumber\\
O_3&=& \bar s \gamma_\mu(1-\gamma_5)b \sum_{q'}
\bar q' \gamma^\mu(1-\gamma_5) q',\nonumber\\
O_4 &=& \bar s_\alpha \gamma_\mu(1-\gamma_5)b_\beta \sum_{q'}
\bar q'_\beta \gamma^\mu(1-\gamma_5) q'_\alpha,\nonumber\\
O_5&=&\bar s \gamma_\mu(1-\gamma_5)b \sum_{q'} \bar q'
\gamma^\mu(1+\gamma_5)q',\nonumber\\
O_6& = &\bar s_\alpha \gamma_\mu(1-\gamma_5)b_\beta \sum_{q'}
\bar q'_\beta \gamma^\mu(1+\gamma_5) q'_\alpha,\nonumber\\
O_7&=& \frac{3}{2}\bar s \gamma_\mu(1-\gamma_5)b \sum_{q'}
e_{q'}\bar q' \gamma^\mu(1+\gamma_5) q',\nonumber\\
O_8 &=&\frac{3}{2} \bar s_\alpha \gamma_\mu(1-\gamma_5)b_\beta \sum_{q'}
e_{q'}\bar q'_\beta \gamma^\mu(1+\gamma_5) q'_\alpha,\nonumber\\
O_9&=&\frac{3}{2}\bar s \gamma_\mu(1-\gamma_5)b \sum_{q'} e_{q'}\bar q'
\gamma^\mu(1-\gamma_5)q',\nonumber\\
O_{10}& = &\frac{3}{2}\bar s_\alpha \gamma_\mu(1-\gamma_5)b_\beta \sum_{q'}
e_{q'}\bar q'_\beta \gamma^\mu(1-\gamma_5) q'_\alpha,
\label{2b}
\vspace{2mm}
\end{eqnarray}
where $\alpha$ and $\beta$ are color indices, the sum index $q^\prime$ runs over the "active" flavors quarks at the scale $m_b$, and $e_{q^\prime}$ is the electric
charge of the quark $q^\prime$($q^\prime=u, d, s, c$ or $b$ quarks). In Eq. (\ref{2b}) $O_1^u$ and $O_2^u$ are tree
operators, $O_3$--$O_6$ are QCD penguin operators and $O_7$--$O_{10}$ are
the operators associated with electroweak penguin diagrams.

The Wilson coefficients, $C_{i}(\mu)$, represent the power contributions
from scales higher than $\mu$ (which refer to the long-distance
contributions)~\cite{Buras:1998raa}. Since the QCD has the property of asymptotic freedom, they can be calculated in perturbation
theory. The Wilson coefficients include the contributions of all heavy particles, such as
the top quark, the $W^{\pm}$ bosons, and so on. Usually, the scale $\mu$ is
chosen to be of order ${\cal O}(m_{b})$ for $B$ meson decays. Since we work in the leading order of perturbative QCD
($O(\alpha_s )$), it is consistent to use the leading order Wilson
coefficients. So, we use numerical values of $C_i(m_b)$ as follow \cite{Keum:2000wi,Lu:2000em}:
\begin{eqnarray}
C_1 &=&-0.2703,\;\; \; \; \; \; \,C_2=1.1188,\nonumber\\
C_3 &=& 0.0126,\;\; \; \; \; \; \; \; \; \;C_4 = -0.0270,\nonumber\\
C_5 &=& 0.0085,\;\; \; \; \; \; \; \; \; \;C_6 = -0.0326,\nonumber\\
C_7 &=& 0.0011,\;\; \; \; \; \; \; \; \; \;C_8 = 0.0004,\nonumber\\
C_9&=& -0.0090,\;\; \; \; \; \; \;C_{10} = 0.0022.
\label{2k}
\vspace{2mm}
\end{eqnarray}
The Wilson coefficients $a_{1}$--$a_{10}$ are defined as usual \cite{Ali:1998eb,Ali:1998gb,Keum:2000ms,Lu:2000hj}:
\begin{eqnarray}
a_1&=&C_2+C_1/3,\;\; \; \; \; \;a_2=C_1+C_2/3,\nonumber \\
a_3&=&C_3+C_4/3,\;\; \; \; \; \;a_4=C_4+C_3/3,\nonumber \\
a_5&=&C_5+C_6/3,\;\; \; \; \; \;a_6=C_6+C_5/3,\nonumber \\
a_7&=&C_7+C_8/3,\;\; \; \; \; \;a_8=C_8+C_7/3,\nonumber \\
a_9&=&C_9+C_{10}/3,\;\; \; \;a_{10}= C_{10}+C_{9}/3.
\label{2ka}
\vspace{2mm}
\end{eqnarray}

For the decay channel of $\bar B_s^0\to M_2M_3$, we denote the emitted meson as $M_2$ while the recoiling meson is
$M_3$. The  $M_2$ ($\rho$ or $\omega$) and the final-state $M_3$ ($\phi$) move along the direction of $n_+=(1,0,{\bf{0}}_T)$ and $n_-=(0,1,{\bf{0}}_T)$ in the light-cone coordinates, respectively. We denote the ratios ${r_\phi } = \frac{{{M_\phi }}}{{{M_{{B_s}}}}}$, ${r_\rho } = \frac{{{M_\rho }}}{{{M_{{B_s}}}}}$ and ${r_\omega}= \frac{{M_\omega}}{{M_{{B_s}}}}$. In the limit ${M_\phi }$, ${M_\rho }$, ${M_\omega}$ $\to 0$, one can  drop the terms of proportional to  $r_\phi ^2$, $r_\rho ^2$, $r_\omega ^2$ safely. The symbols $P_{B}$, $P_2$ and $P_3$ refer to the $\bar B_s$ meson momentum, the $\rho(\omega)$ meson momentum, and the final-state $\phi$ momentum, respectively.
Under the above approximation, the momenta can be written as:
\begin{eqnarray}
P_B&=&\frac{M_{B_s}}{\sqrt 2}(1,1,{\bf{0}}_T),\;\;
P_2=\frac{M_{B_s}}{\sqrt 2}(1,0,{\bf{0}}_T),\;\;
P_3=\frac{M_{B_s}}{\sqrt 2}(0,1, {\bf{0}}_T).
\end{eqnarray}
One can denote the light (anti-)quark momenta $k_1$, $k_2$ and $k_3$ for the mesons $B_s$, $\rho(\omega)$ and $\phi$, respectively. We can write:
\begin{eqnarray}
k_1&=&(x_1\frac{M_{B_s}}{\sqrt 2},0, {\bf k}_{1\perp}),\;\;
k_2=(x_2\frac{M_{B_s}}{\sqrt 2},0, {\bf k}_{2\perp}),\;\;
k_3=(0,x_3\frac{M_{B_s}}{\sqrt 2},{\bf k}_{3\perp}),
\end{eqnarray}
where $x_1$, $x_2$ and $x_3$ are the momentum fraction. ${\bf k}_{1\perp}$, ${\bf k}_{2\perp}$ and ${\bf k}_{3\perp}$ refer to the transverse momentum
of the quark, respectively.  The longitudinal polarization vectors of the $\rho(\omega)$ and $\phi$ are given as
\begin{eqnarray}
\epsilon_{2}(L)=\frac{P_{2}}{M_{\rho(\omega)}}-\frac{M_{\rho(\omega)}}{P_{2} \cdot n_{-}} n_{-},\;\;\;\; \epsilon_{3}(L)=\frac{P_{3}}{M_{\phi}}-\frac{M_{\phi}}{P_{3} \cdot n_{+}} n_{+},
\end{eqnarray}
which satisfy the orthogonality relationship of $\epsilon_{2}(L) \cdot P_{2}=\epsilon_{3}(L) \cdot P_{3}=0$, and the normalization of $\epsilon_{2}^{2}(L)=\epsilon_{3}^{2}(L)=-1$. The transverse polarization vectors can be adopted diredtly as
\begin{equation}
\epsilon_{2}(T)=\left(0,0,{\bf{1}}_{T}\right), \quad \epsilon_{3}(T)=\left(0,0,{\bf{1}}_{T}\right).
\end{equation}

The wave function of $B_s$ meson can be expressed as
\begin{equation}
 \phi_{B_s}= \frac{i}{\sqrt{6}} (\not \! P_{B_s} +M_{B_s}) \gamma_5
\phi_{B_s} ({ k_1}), \label{bmeson}
\end{equation}
where the distribution amplitude $\phi_{B_s}$ is shown in Refs. \cite{Ali:2007ff,Li:2004ep,Wang:2014mua}:
\begin{equation}
\phi_{B_s}(x,b) = N_{B_s} x^2(1-x)^2 \exp \left[ -\frac{M_{B_s}^2\
x^2}{2 \omega_b^2} -\frac{1}{2} (\omega_b b)^2 \right].\label{waveb}
\end{equation}
The shape parameter $\omega_b$ is a free parameter.
Based on lattice QCD and the light-cone sum rule \cite{Li:2003yj},
we take $\omega_b=0.50~\mathrm{GeV}$ for the $B_s$ meson.
The normalization factor $N_{B_s}$ depends on the values
of $\omega_b$ and the decay constant $f_{B_s}$, which is defined through the
normalization relation $\int_0^1 {dx{\phi _{{B_s}}}\left( {x,0} \right)}  = {f_{{B_s}}}/(2\sqrt 6)$.

The distribution amplitudes of vector meson(V=$\rho$, $\omega$ or $\phi$), $\phi_{V}$, $\phi_{V}^T$, $\phi^t_{V}$, $\phi^s_{V}$, $\phi^v_{V}$, and $\phi^a_{V}$,
are calculated using light-cone QCD sum rule \cite{Ball:1998ff,Ball:2004rg}:
\begin{eqnarray}
\phi_\rho (x)&=&\frac{3f_\rho}{\sqrt{6}} x (1-x)\left[1+0.15C_2^{3/2} (t) \right]\;,\label{phirho}\\
\phi_\omega(x)&=&\frac{3f_\omega}{\sqrt{6}} x (1-x)\left[1+0.15C_2^{3/2} (t)\right]\;,\label{phiomega}\\
\phi_{\phi}(x)&=&\frac{3f_{\phi}}{\sqrt{6}} x (1-x)\left[1+0.18C_2^{3/2} (t)\right]\;,\label{phiphi} \\
\phi_V^T(x)&=&\frac{3f_V^T}{\sqrt{6}} x (1-x)\left[1+0.14C_2^{3/2} (t) \right]\;,\label{phirho1}\\
\phi^t_V(x)&=&\frac{3f^T_V}{2\sqrt 6}t^2\;,\label{phitv}\\
\phi^s_V(x)&=&\frac{3f_V^T}{2\sqrt 6} (-t)\;,\label{phisv}\\
\phi_V^v(x)&=&\frac{3f_V}{8\sqrt6}(1+t^2)\;,\label{phivv}\\
 \phi_V^a(x)&=&\frac{3f_V}{4\sqrt6}(-t)\;,\label{phiav}
\end{eqnarray}
where $t=2x-1$. Here $f_{V}$ is the decay constant of
the vector meson with longitudinal  polarization. The Gegenbauer polynomials $C^{\nu}_{n}(t)$ can be
found easily in Refs. \cite{Fan:2012kn,Huang:2005if}.

\section{\label{sec:cpv1}$CP$ violation in $\bar{B}_s^0\to\rho(\omega )\phi\to {\pi^+}{\pi^-}\phi$ decay process}
\subsection{\label{subsec:form}Formalism}

The hadronic decay rare for the process of $\bar{B}_s^0\to\rho(\omega ) \phi$ is written as 
\begin{equation}
\Gamma  = \frac{{{P_c}}}{{8\pi M_{{B_s}}^2}}\sum\limits_{\sigma=L,T}{{A^{(\sigma) \dagger }}{A^{(\sigma )}}}, \label{dr1}
\end{equation}
where $P_c=|P_{2z}|=|P_{3z}|$ is the momentum of the vector meson. The superscript $\sigma$ denotes the
helicity states of the two vector mesons with the longitudinal (transverse) components L(T). The amplitude
$A^{(\sigma)}$ is decomposed into \cite{Li:2004ti,Huang:2005if,Lu:2005be}
\begin{eqnarray}
A^{(\sigma)}=M^2_{B_{s}}A_{L}+M^2_{B_{s}}A_{N}
\epsilon^{*}_{2}(\sigma=T)\cdot\epsilon^{*}_{3}(\sigma=T) +i
A_{T}\epsilon^{\alpha \beta\gamma \rho}
\epsilon^{*}_{2\alpha}(\sigma)\epsilon^{*}_{3\beta}(\sigma)
P_{2\gamma }P_{3\rho }\;,
\end{eqnarray}
with the convention $\epsilon^{0123}=1$. The amplitude $A_{i}$
($i$ refer to the three kind of polarizations, longitudinal (L), normal (N) and transverse (T)) can be written as
\begin{eqnarray}
M^2_{B_{s}}A_L &=& a \,\, \epsilon_2^{*}(L) \cdot \epsilon_3^{*}(L) +{\frac{b}{M_2 M_{3}}} \epsilon_{2}^{*}(L) \cdot P_3 \,\, \epsilon_{3}^{*}(L) \cdot P_2\;, \nonumber \\
M^2_{B_{s}}A_N &=& a \;, \label{id-rel}  \nonumber \\
A_T &=& {\frac{c}{M_2 M_{3}}}\;,
\end{eqnarray}
where $a$, $b$ and $c$ are the Lorentz-invariant amplitudes. $M_2$, $M_3$ refer to the masses of the vector mesons $\rho(\omega)$ and $\phi$, respectively.

The longitudinal $H_{0}$, transverse $H_{\pm}$ of helicity amplitudes can be expressed
\begin{eqnarray}
H_{0}&=&M^2_{B_{s}}A_{L}, \nonumber\\
H_{\pm}&=&M^2_{B_{s}}A_{N}\mp M_{2}M_{3} \sqrt{\kappa^2-1}A_{T},
\label{dr1bbbbbbbbb}
\end{eqnarray}
where $H_{0}$, $H_{+}$ and $H_{-}$ are the tree-level and penguin-level helicity amplitudes of the decay process $\bar B_s^0 \to \rho (\omega )\phi  \to {\pi ^ + }{\pi ^ - }\phi $ from the three kind of polarizations, respectively. The helicity summation satisfy the relation,
\begin{equation}
\sum\limits_{\sigma=L,R}{{A^{(\sigma) \dagger }}{A^{(\sigma )}}}=|H_{0}|^{2}+|H_{+}|^{2}+|H_{-}|^{2}. \label{dr1aaa}
\end{equation}

In the VMD model \cite{Nambu:1997vw,Sakurai:1969zz}, the vacuum polarisation of the photons are assumed to be coupled through the vector meson ($\rho$ meson). 
Based on the same mechanism, $\rho-\omega$ mixing was proposed and later gradually applied to $B$ meson physics.
The formalism for the $CP$ violation in $B$ hadronic decays can be generalized to $B_s$ in a straightforward manner \cite{Enomoto:1996cv,Gardner:1997yx,Guo:2000uc}. According to the effective Hamiltonian, the amplitude $A$ ($\bar{A}$) for the decay process of
$\bar B_s^0 \to  {\pi ^ + }{\pi ^ - } \phi$
($B_s^0 \to {\pi ^ + }{\pi ^ - } \bar \phi$) can be written as \cite{Gardner:1997yx}:
\begin{eqnarray}
A&=&\big<\pi^+\pi^{-}\phi|H^T|\bar{B}_{s}^{0}\big>+\big<\pi^+\pi^{-}\phi|H^P|\bar{B}_{s}^{0}\big>,\label{A}\\
\bar{A}&=&\big<\pi^+\pi^{-}\bar\phi|H^T|{B}_{s}^{0}\big>+\big<\pi^+\pi^-\bar\phi|H^P|{B}_{s}^{0}\big>,\label{Abar}
\end{eqnarray}
where $H^T$ and $H^P$ are the Hamiltonian for the tree and penguin operators, respectively.

The relative magnitude and phases between the tree
and penguin operator contribution are defined as follows:
\begin{eqnarray}
A&=&\big<\pi^+\pi^{-}\phi|H^T|\bar{B}_{s}^{0}\big>[1+re^{i(\delta+\phi)}],\label{A'}\\
\bar{A}&=&\big<\pi^+\pi^-\bar\phi|H^T|{B}_{s}^{0}\big>[1+re^{i(\delta-\phi)}],
\label{A'bar}
\end{eqnarray}
where $\delta$ and $\phi$ are strong and weak phases, respectively.
The weak phase difference $\phi$ can be expressed as a combination of the CKM matrix elements, and it is
$\phi=\arg[(V_{tb}V_{ts}^{*})/(V_{ub}V_{us}^{*})]$ for the $b \to s$ transition. The parameter $r$ is the absolute value of the ratio of tree and penguin amplitudes:
\begin{eqnarray}
r\equiv\Bigg|\frac{\big<\pi^+\pi^{-}\phi|H^P|\bar{B}_{s}^{0}\big>}{\big<\pi^+\pi^{-}\phi|H^T|\bar{B}_{s}^{0}\big>}\Bigg|
\label{r}.
\end{eqnarray}
The parameter of $CP$ violating asymmetry, $A_{CP}$, can be written as
\begin{eqnarray}
A_{CP}=\frac{|A|^{2}-|\bar{A}|^{2}}{|A|^{2}+|\bar{A}|^{2}} =\frac{-2(T_{0}^2r_{0}\sin\delta_0+T_{+}^2r_{+}\sin\delta_{+}+T_{-}^2r_{-}\sin\delta_-)\sin\phi}  {\sum_{i=0+-}T_{i}^2(1+r_{i}^2+2r_{i}\cos\delta_i\cos\phi)},
\label{eq:CP-tuidao}
\end{eqnarray}
where $T_{i}(i=0,+,-)$ are the tree-level helicity amplitudes of the decay process $\bar B_s^0 \to  {\pi ^ + }{\pi ^ - } \phi$ from $H_{0}$, $H_{+}$ and $H_{-}$ of the Eq. (\ref{dr1bbbbbbbbb}), respectively. $r_{j}(j=0,+,-)$ refer to the absolute value of the ratio of tree and penguin amplitude for the three kind of polarizations, respectively.
$\delta_{k}(k=0,+,-)$ represent the relative strong phases between the tree
and penguin operator contributions from three kind of helicity amplitudes.
We can see explicitly from Eq. (\ref{eq:CP-tuidao}) that both weak and strong phase
differences are responsible for $CP$ violation. In order to obtain a large signal for direct CP violation, we
need some mechanism to change either $\sin\delta$ or $r$. 
With this mechanism, working at the first order of isospin violation, we have the following results
when the invariant mass of $\pi^+\pi^-$ is near the $\omega$ resonance mass \cite{Gardner:1997yx,Guo:1998eg}:
\begin{eqnarray}
\big<\pi^+\pi^-\phi|H^T|\bar{B}^0_{s}\big>=\frac{g_{\rho}}{s_{\rho}s_{\omega}}\widetilde{\Pi}_{\rho\omega}t_{\omega}^{i}+\frac{g_{\rho}}{s_{\rho}}t_{\rho}^{i},
\label{Htr}\\
\big<\pi^+\pi^-\phi|H^P|\bar{B}^0_{s}\big>=\frac{g_{\rho}}{s_{\rho}s_{\omega}}\widetilde{\Pi}_{\rho\omega}p_{\omega}^{i}+\frac{g_{\rho}}{s_{\rho}}p_{\rho}^{i},
\label{Hpe}
\end{eqnarray}
where $t_{\rho}^{i}(p_{\rho}^{i})$ and $t_{\omega}^{i}(p_{\omega}^{i})$ are the tree (penguin)-level helicity amplitudes
for $\bar{B}_{s}\rightarrow\rho^0\phi$ and
$\bar{B}_{s}\rightarrow\omega \phi$, respectively.The amplitudes $t_{\rho}^{i}$, $p_{\rho}^{i}$, $t_{\omega}^{i}$ and $p_{\omega}^{i}$ can be
found  in Sec. \ref{cal}. 
$g_{\rho}$ is
the coupling for $\rho^0\rightarrow\pi^+\pi^-$.
$\widetilde{\Pi}_{\rho\omega}$ is the effective $\rho-\omega$
mixing amplitude which also effectively includes the direct
coupling $\omega\rightarrow\pi^+\pi^-$ \cite{OConnell:1997ggd}. $s_{V}$, $m_{V}$ and $\Gamma_V$($V$=$\rho$ or
$\omega$) is the inverse propagator, mass and decay rate of the vector meson $V$, respectively. $s_V$ can be expressed as
\begin{eqnarray}
s_V=s-m_V^2+{\rm{i}}m_V\Gamma_V,
\end{eqnarray}
with $\sqrt{s}$ being the invariant masses of the $\pi^+\pi^-$ pairs. The numerical values for the $\rho- \omega$ mixing parameter $\widetilde{\Pi}_{\rho\omega}(s)={\rm{Re}}\widetilde{\Pi}_{\rho\omega}(m_{\omega}^2)+{\rm{Im}} \widetilde{\Pi}_{\rho\omega}(m_{\omega}^2)$ are \cite{Lu:2018fqe}
\begin{eqnarray}
{\rm{Re}} \widetilde{\Pi}_{\rho\omega}(m_{\omega}^2)&=&-4760\pm{440}\,
\rm{MeV}^2,\nonumber\\ {\rm{Im}} \widetilde{\Pi}_{\rho\omega}(m_{\omega}^2)&=&-6180\pm{3300}\,
\textrm{MeV}^2. \label{rhoomegamixing}
\end{eqnarray}
From Eqs. (\ref{A}), (\ref{A'}), (\ref{Htr}) and (\ref{Hpe}) one has
\begin{eqnarray}
re^{i\delta_{i}}e^{i\phi}=\frac{\widetilde{\Pi}_{\rho\omega}p_{\omega}^{i}+s_{\omega}p_{\rho}^{i}}{\widetilde{\Pi}_{\rho\omega}t_{\omega}^{i}+s_{\omega}t_{\rho}^{i}},
\label{rdtdirive}
\end{eqnarray}
Defining \cite{Enomoto:1996cv,Enomoto:1997bq}
\begin{eqnarray}
\frac{p_{\omega}^{i}}{t_{\rho}^{i}}\equiv r^\prime
e^{i(\delta^{i}_q+\phi)},\quad\frac{t_{\omega}^{i}}{t_{\rho}^{i}}\equiv
\alpha
e^{i\delta^{i}_\alpha},\quad\frac{p_{\rho}^{i}}{p_{\omega}^{i}}\equiv
\beta e^{i\delta^{i}_\beta}, \label{def}
\end{eqnarray}
where $\delta^{i}_\alpha$, $\delta^{i}_\beta$ and $\delta^{i}_q$ are strong
phases form the three kind of polarizations, respectively. One finds the following expression from Eqs. (\ref{rdtdirive}) and (\ref{def}):
\begin{eqnarray}
re^{i\delta_{i}}=r^\prime
 e^{i\delta^{i}_q}\frac{\widetilde{\Pi}_{\rho\omega}+\beta
e^{i\delta^{i}_\beta}s_{\omega}}{\widetilde{\Pi}_{\rho\omega}\alpha
e^{i\delta^{i}_\alpha}+s_{\omega}}. \label{rdt}
\end{eqnarray}
$\alpha e^{i\delta^{i}_\alpha}$, $\beta e^{i\delta^{i}_\beta}$, and $r^\prime e^{i \delta^{i}_q}$ will be calculated later. In order to obtain the $CP$ violating asymmetry in Eq.
(\ref{eq:CP-tuidao}), $A_{cp}$, sin$\phi$ and cos$\phi$ are needed, where $\phi$ is
determined by the CKM matrix elements. In the Wolfenstein
parametrization \cite{Wolfenstein:1964ks}, the weak phase $\phi$ comes from $[{V_{tb}}V_{ts}^*/{V_{ub}}V_{us}^*]$.
One has
\begin{eqnarray}
{\rm sin}\phi &=&-\frac{\eta }{\sqrt{\rho ^2+\eta ^2}}, \nonumber \\
{\rm cos}\phi &=&-\frac{\rho }{\sqrt{\rho ^2+\eta ^2}},
\label{3l1}
\vspace{2mm}
\end{eqnarray}
where the same result has been used for $b \to s$ transition from Ref. \cite{Guo:1999ip}.

\subsection{\label{cal}Calculation details}

We can decompose the decay amplitude for the decay process $\bar{B}_{s}^0\rightarrow \rho^{0}(\omega)\phi$
in terms of tree-level and penguin-level contributions depending on the CKM matrix elements of $V_{ub}V^{*}_{us}$ and $V_{tb}V^{*}_{ts}$.
From Eqs. (\ref{eq:CP-tuidao}), (\ref{rdtdirive}) and (\ref{def}), in order to obtain the formulas of the $CP$ violation, we need calculate the amplitudes $t_{\rho}$, $t_{\omega}$, $p_{\rho}$ and $p_{\omega}$ in perturbative QCD approach.
The $F$ and $M$ function can be found in the Appendix from the perturbative QCD approach.

\begin{figure}[tbh]
	\centering
	\includegraphics[width=0.23\textwidth]{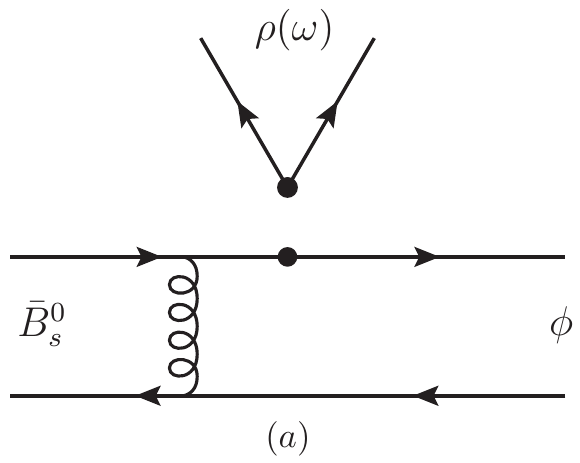}
	\hspace{0.2in}
	\includegraphics[width=0.23\textwidth]{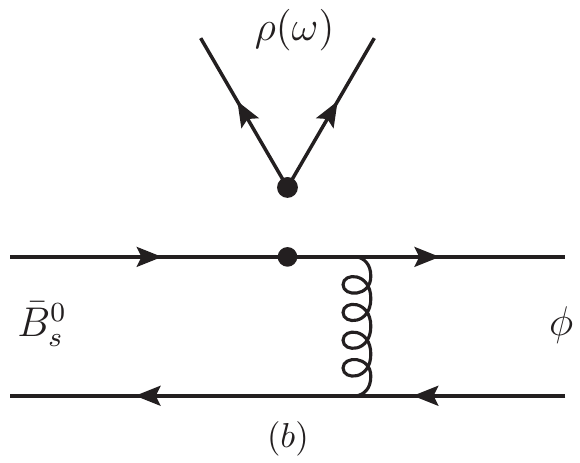}	\\  \vspace{1em}
	\includegraphics[width=0.23\textwidth]{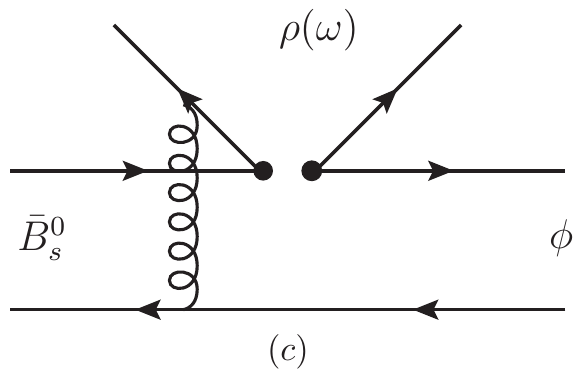}
	\hspace{0.2in}
	\includegraphics[width=0.23\textwidth]{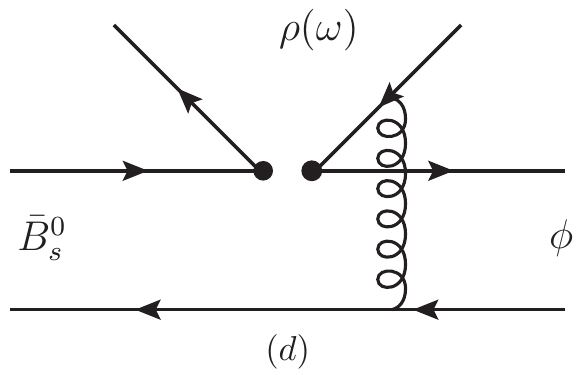}	
\caption{\label{emission plot} Leading order Feynman diagrams for $\bar{B}_s^0 \to \rho(\omega)\phi$} 
\end{figure}

There are four types of Feynman diagrams contributing to $\bar{B}_{s}^0\rightarrow \rho^{0}(\omega)\phi$ emission decay mode. The leading order diagrams in pQCD approach are shown in Fig.\ref{emission plot}. The first two diagrams in Fig.\ref{emission plot} (a)(b) are called factorizable diagrams and the last two diagrams in Fig.\ref{emission plot} (c)(d) are called non-factorizable diagrams \cite{Zhu:2005rx,Chen:2002pz}.
The relevant decay amplitudes can be easily obtained by these hard gluon exchange diagrams and the Lorenz structures of the mesons wave functions.
Through calculating these diagrams, the formulas of $\bar{B}_{s}^0 \to \rho \phi$ or $\bar{B}_{s}^0 \to \omega \phi$ are similar to those of $B\to \phi K^*$ and $B_s \to K^{*-}K^{*+}$ \cite{Chen:2002pz,Zou:2015iwa}.
We just need to replace some corresponding wave functions, Wilson coefficients and corresponding parameters.

With the Hamiltonian Eq. (\ref{2a}), depending on CKM matrix elements of $V_{ub}V^{*}_{us}$ and  $V_{tb}V^{*}_{ts}$,
the electroweak penguin dominant decay amplitudes $A^{(i)}$ for $\bar B_{s}^0\rightarrow \rho^{0}\phi$ in pQCD can be
written as
\begin{eqnarray}
\sqrt{2}A^{(i)}(\bar B_{s}^0\to\rho^{0}\phi)&=&V_{ub}V_{us}^{*}T_{\rho}^{i}-V_{tb}V_{ts}^{*}P_{\rho}^{i} , \label{BcDrho1}
\end{eqnarray}
where the superscript $i$ denote different helicity
amplitudes $L,N$ and $T$. The longitudinal $t_{\rho(\omega) }^0$, transverse $t_{\rho(\omega) }^{\pm}$ of helicity amplitudes satisfy relationship of $t_{\rho(\omega) }^0= t_{\rho(\omega) }^L$ and $t_{\rho(\omega) }^ \pm  = ({{t_{\rho(\omega) }^N \mp t_{\rho(\omega) }^T}})/({{\sqrt 2 }})$. The amplitudes from tree and penguin diagrams can be written as 
 $T_\rho ^i = t_\rho ^i/{V_{ub}}V_{us}^*$ and $P_\rho ^i = p_\rho ^i/{V_{tb}}V_{ts}^*$, respectively. 
The formula for the tree level amplitude is
\begin{eqnarray}
T_{\rho}^{i}=\frac{G_F}{\sqrt{2}}\Big \{ f_{\rho} F_{
B_s\to \phi}^{LL,i} \left[a_{2}\right]+ M_{ B_s\to
\phi}^{LL,i}\left[C_{2}\right]\Big \},  \label{trho1}
\end{eqnarray}
where $f_{\rho}$ refers to the decay constant of $\rho$ meson. The penguin level amplitude are expressed in the following
\begin{eqnarray}   
P_{\rho}^{i}=\frac{G_F}{\sqrt{2}} \Big \{
f_{\rho}F_{B_s\to \phi}^{LL,i}\left[\frac{3}{2}(a_{9}+a_{7})
\right] +M_{ B_s\to \phi}^{LL,i} \left[\frac{3}{2}C_{10}\right]
-M_{B_s\to \phi}^{SP,i}\left[\frac{3}{2}C_{8}\right] \Big\}~.  \label{Prho1}
\end{eqnarray}
The QCD penguin dominant decay amplitude for $\bar B_{s}^0\to \omega \phi$ can be written as
\begin{eqnarray}
\sqrt 2A^i(\bar B_{s}^0\to \omega\phi)= V_{ub}V_{us}^{*} T_{\omega}^{i} -  V_{tb}V_{ts}^{*} P_{\omega}^{i}, \label{BcDomega1}
\end{eqnarray}
where $T_\omega ^i = t_\omega ^i/{V_{ub}}V_{us}^*$ and $P_\omega ^i = p_\omega ^i/{V_{tb}}V_{ts}^*$ which refer to the tree and penguin amplitude, respectively.
We can give the tree level contribution in the following
\begin{eqnarray}
T_{\omega}^{i}=\frac{G_F}{\sqrt{2}}\Big\{f_\omega F_{B_s\to
\phi}^{LL,i}\left[a_{2}\right] + M_{B_s\to
\phi}^{LL,i}\left[C_{2}\right]\Big \}, \label{tomega1}
\end{eqnarray}
where $f_{\omega}$ refers to the decay constant of $\omega$ meson.
The penguin level contribution are given as following
\begin{eqnarray}
P_{\omega}^{i}&=& \frac{G_F}{\sqrt{2}} \Big\{ f_\omega F_{B_s\to \phi}^{LL,i}\left[2a_3
+ 2a_5 +\frac{1}{2}a_7+\frac{1}{2}a_9
   \right]\nonumber
   \\
&&+ M_{B_s\to \phi}^{LL,i} \left[2C_4+\frac{1}{2}C_{10}\right]
-M_{B_s\to \phi}^{SP,i}\left[2C_6+\frac{1}{2}C_8\right]\Big \}. \label{pomega1}
\end{eqnarray}
Based on the definition of Eq. (\ref{def}), we can get
\begin{eqnarray}
\alpha e^{i\delta^{i}_\alpha}&=&\frac{t_{\omega}^{i}}{t_{\rho}^{i}}, \label{eq:afaform} \\
\beta e^{i\delta^{i}_\beta}&=&\frac{p_{\rho}^{i}}{p_{\omega}^{i}}, \label{eq:btaform}\\
r^\prime e^{i\delta^{i}_q}&=&\frac{P^{i}_{\omega}}{T^{i}_{\rho}}
\times\bigg|\frac{V_{tb}V_{ts}^*}{V_{ub}V_{us}^*}\bigg|,  \label{eq:delform}
\end{eqnarray}
where
\begin{equation}
\left|\frac{V_{tb}V^{*}_{ts}}{V_{ub}V^{*}_{us}}\right|=\frac{\sqrt{\rho^2+\eta^2}}{\lambda^2(\rho^2+\eta^2)}.
\label{3p}
\vspace{2mm}
\end{equation}
From above equations, the new strong phases ${\delta^{i} _\alpha }$, $\delta^{i}_\beta$ and $\delta^{i}_q$ are obtained from tree and penguin diagram contributions by the $\rho-\omega$ interference. The total
strong phase $\delta_{i}$ are obtained by the Eqs. (\ref{def}) and (\ref{rdt}) in the framework of pQCD.

\section{\label{int}Input parameters}

The CKM matrix, which elements are determined from experiments, can be expressed in terms of the Wolfenstein parameters $A$, $\rho$, $\lambda$ and $\eta$ \cite{Wolfenstein:1964ks,Tanabashi:2018oca}:
\begin{equation}
\left(
\begin{array}{ccc}
  1-\frac{1}{2}\lambda^2   & \lambda                  &A\lambda^3(\rho-\mathrm{i}\eta) \\
  -\lambda                 & 1-\frac{1}{2}\lambda^2   &A\lambda^2 \\
  A\lambda^3(1-\rho-\mathrm{i}\eta) & -A\lambda^2              &1\\
\end{array}
\right),\label{ckm}
\end{equation}
where $\mathcal{O} (\lambda^{4})$ corrections are neglected. The latest values for the parameters in the CKM matrix are \cite{Tanabashi:2018oca}:
\begin{eqnarray}
&& \lambda=0.22453\pm0.00044,\quad A=0.836\pm0.015,\nonumber \\
&& \bar{\rho}=0.122_{-0.017}^{+0.018},\quad
\bar{\eta}=0.355_{-0.011}^{+0.012}.\label{eq: rhobarvalue}
\end{eqnarray}
where
\begin{eqnarray}
 \bar{\rho}=\rho(1-\frac{\lambda^2}{2}),\quad
\bar{\eta}=\eta(1-\frac{\lambda^2}{2}).\label{eq: rho rhobar
relation}
\end{eqnarray}
From Eqs. (\ref{eq: rhobarvalue}) and (\ref{eq: rho rhobar relation})
we have
\begin{eqnarray}
0.108<\rho<0.144,\quad  0.353<\eta<0.377.\label{eq: rho value}
\end{eqnarray}
%
The other parameters and the corresponding references are listed in Table \ref{table2}.
\begin{table*}[t]
\setlength{\abovecaptionskip}{0pt}
\setlength{\belowcaptionskip}{8pt}
\begin{center}
\renewcommand\arraystretch{1.5}
\tabcolsep 0.25in
\caption{Input parameters} \label{table2}
\begin{tabular}{lll}
\hline \hline
Parameters&Input data & References  \\ \hline
Fermi constant (in $\rm{GeV}^{-2}$)&$G_F=1.16638\times10^{-5}.$& \cite{Tanabashi:2018oca}\\
                        &$m_{B^0_s}=5366.89,~\tau_{B^0_s}=1.509\times10^{-12}s,$& \\
                        &$m_{\rho^0(770)}=775.26, ~\Gamma_{\rho^0(770)}=149.1,$&\\
Masses and decay widths (in MeV)  &$m_{\omega(782)}=782.65, ~\Gamma_{\omega(782)}=8.49,$& \cite{Tanabashi:2018oca}\\
                        &$m_\pi=139.57,~m_\phi=1019.461.$&\\
                       &$f_\rho=215.6\pm5.9,~f_\rho^T=165\pm9,$&\\
Decay constants (in MeV)   &$f_\omega=196.5\pm4.8,~f_\omega^T=145\pm10,$& \cite{Straub:2015ica,Liu:2016rqu}\\
               &$f_\phi=231\pm4,~f_\phi^T=200\pm10.$& \\ \hline \hline
\end{tabular}
\end{center}
\end{table*}
\section{\label{num}The numerical results of $CP$ violation in $\bar{B}^0_{s}\rightarrow \rho^0(\omega)\phi\rightarrow\pi^+\pi^-\phi$}

\begin{figure}[h]
  \centering
\includegraphics[width=0.5\textwidth]{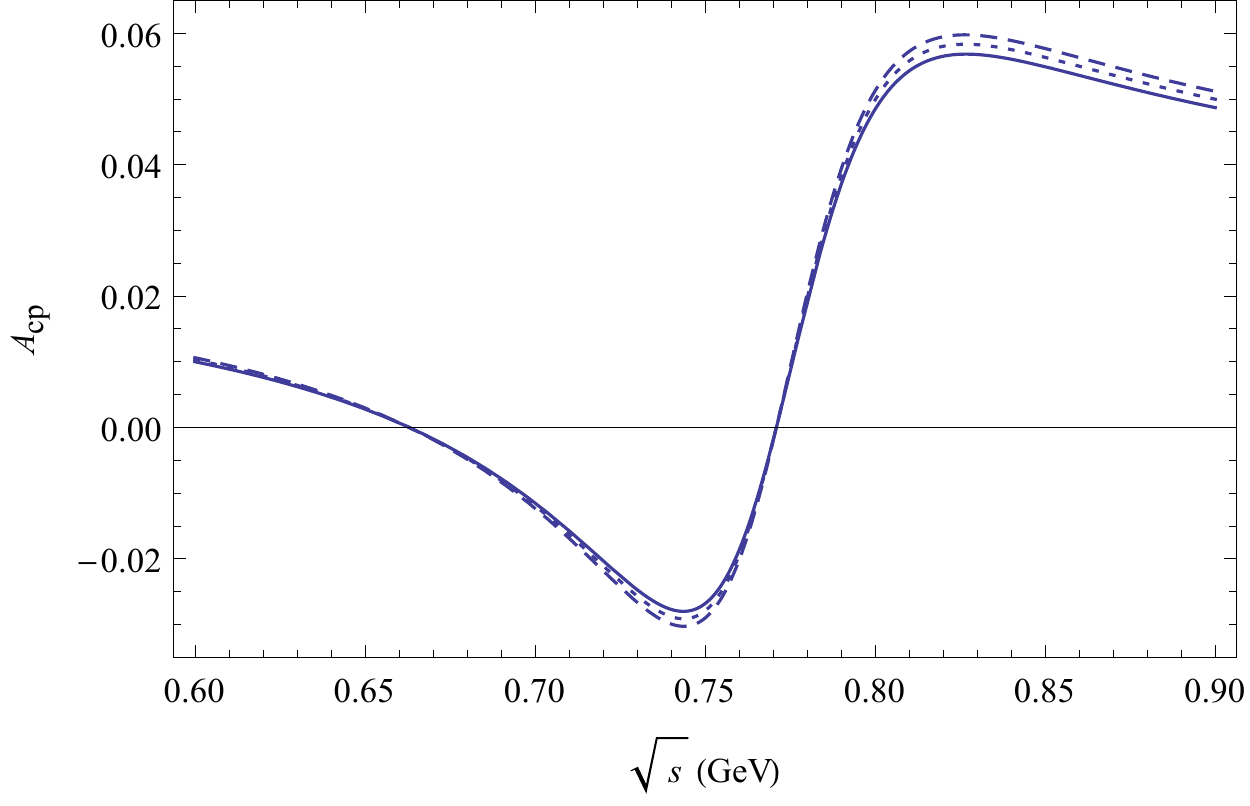}
\caption{\label{Acp plot} The $CP$ violating asymmetry, $A_{cp}$, as a function of $\sqrt{s}$ for
different CKM matrix elements. The dash line, dot line and solid line corresponds to the maximum, middle, and minimum CKM matrix element for the decay channel of $\bar{B}^0_{s}\rightarrow \rho^0(\omega)\phi\rightarrow\pi^+\pi^-\phi$, respectively.}
\end{figure}

We have investigated the $CP$ violating asymmetry, $A_{CP}$, for the $\bar{B}^0_{s}\rightarrow \rho^0(\omega)\phi\rightarrow\pi^+\pi^-\phi$ decay process.  The numerical results of the $CP$ violating asymmetry are shown for the decay process in Fig.~\ref{Acp plot}. It is found that the $CP$ violation can be enhanced via $\rho-\omega$ mixing for the decay channel $\bar{B}^0_{s}\rightarrow \rho^0(\omega)\phi\rightarrow\pi^+\pi^-\phi$ when the invariant mass of $\pi^{+}\pi^{-}$ is in the vicinity of the $\omega$ resonance within perturbative QCD scheme.

\begin{figure}
	\centering
\includegraphics[width=0.5\textwidth]{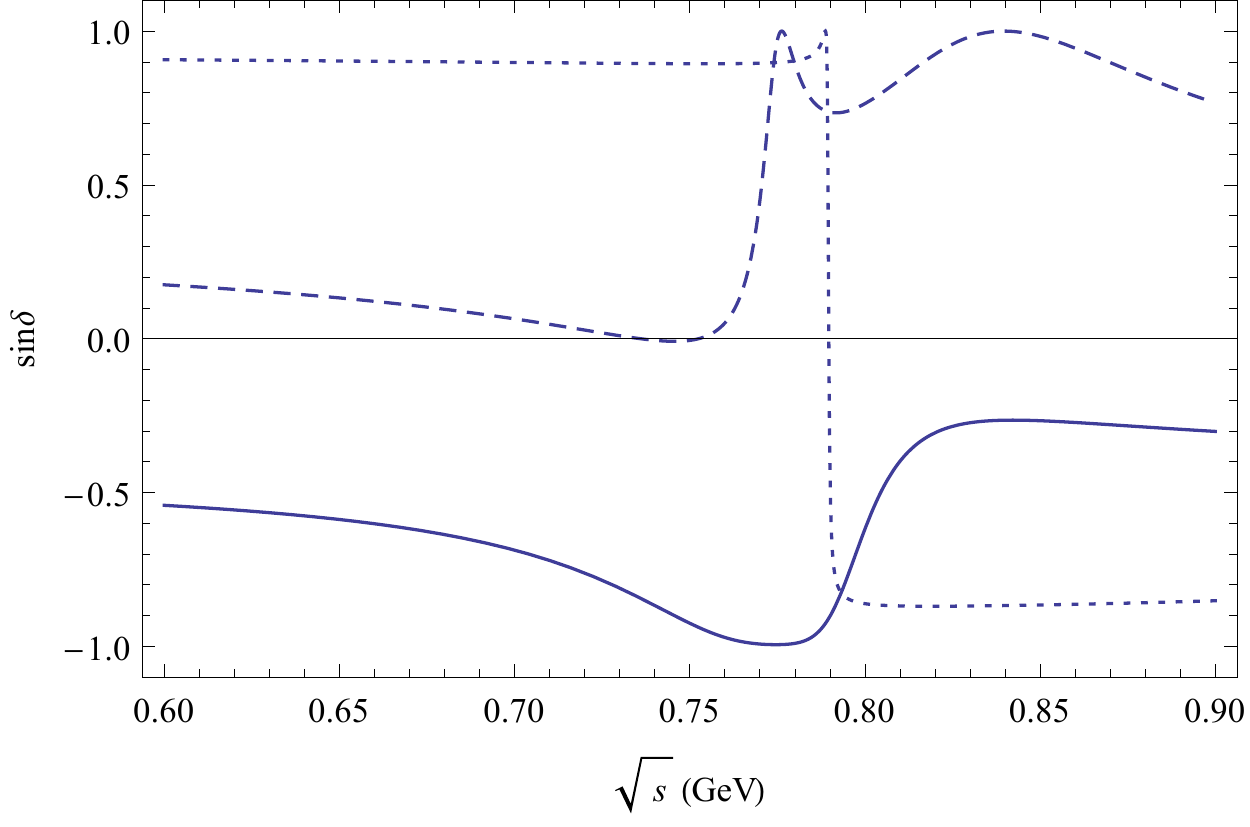}
\caption{\label{sin1 plot} $\sin\delta$ as a function of $\sqrt{s}$
corresponding to central parameter values of CKM matrix elements
for $\bar{B}^0_{s}\rightarrow \rho^0(\omega)\phi\rightarrow\pi^+\pi^-\phi$.
The dash line, dot line and solid line corresponds to $\sin{\delta_{0}}$, $\sin{\delta_{+}}$ and $\sin{\delta_{-}}$, respectively.}
\end{figure}

\begin{figure}
	\centering
\includegraphics[width=0.5\textwidth]{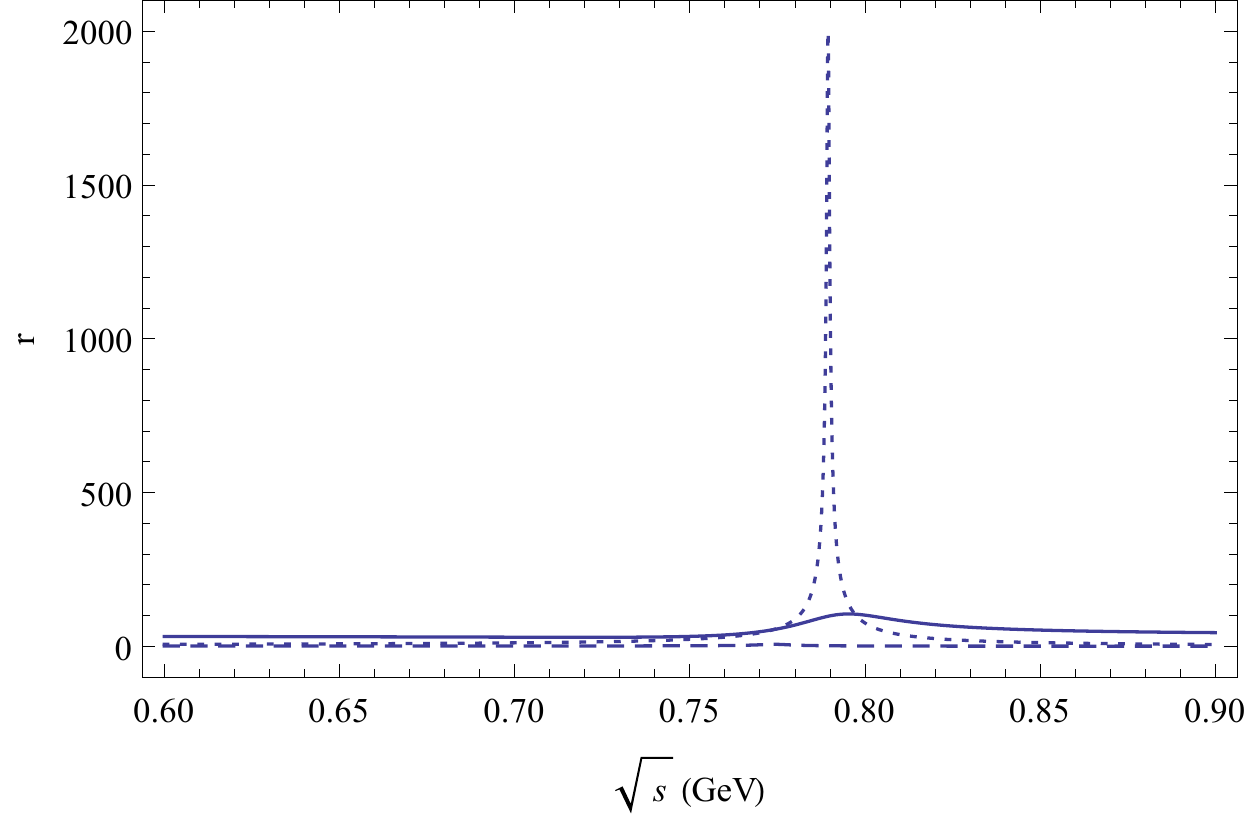}
\caption{\label{r plot} Plot of $r$ as a function of $\sqrt{s}$
corresponding to central parameter values of CKM matrix elements
for $\bar{B}^0_{s}\rightarrow \rho^0(\omega)\phi\rightarrow\pi^+\pi^-\phi$.
The dash line, dot line and solid line corresponds to $r_{0}$, $r_{+}$ and $r_{-}$, respectively.}
\end{figure}

The $CP$ violation depends on the weak phase difference from CKM matrix elements and the strong phase difference. The CKM matrix elements, which relate to $\rho$, $\eta$ and $\lambda$, are given in Eq. (\ref{eq: rhobarvalue}). The uncertainties due to the CKM matrix elements are mostly from $\rho$ and $\eta$ since $\lambda$ is well determined. Hence  we take the central value of $\lambda=0.224$ in Eq. (\ref{eq: rho value}). In our numerical calculations, we let $\rho$, $\eta$ and $\lambda=0.224$ vary among the limiting values. The numerical results are shown from Fig.~\ref{Acp plot} to Fig.~\ref{r plot} with the different parameter values of CKM matrix elements. The dash line, dot line and solid line corresponds to the maximum, middle, and minimum CKM matrix element for the decay channel of $\bar{B}^0_{s}\rightarrow \rho^0(\omega)\phi\rightarrow\pi^+\pi^-\phi$, respectively. We find the $CP$ violation is not sensitive to the CKM matrix elements for the different values of $\rho$ and $\eta$. In Fig.~\ref{Acp plot}, we give the plot of $CP$ violating asymmetry as a function of $\sqrt{s}$. From the Fig.~\ref{Acp plot}, one can see the $CP$ violation parameter is dependent on $\sqrt{s}$ and changes rapidly due to $\rho-\omega$ mixing when the invariant mass of $\pi^{+}\pi^{-}$ is in the vicinity of the $\omega$ resonance. From the numerical results, it is found that the maximum $CP$ violating parameter reaches $5.98\%$  for the decay channel of $\bar{B}^0_{s}\rightarrow\pi^+\pi^-\phi$ in the case of ($\rho_{max}$, $\eta_{max}$).

From Eq. (\ref{eq:CP-tuidao}), one can see that the $CP$ violating parameter is related to sin$\delta$ and $r$. The plots of
$\sin\delta_0$ ($\sin\delta_+$ and $\sin\delta_-$) and $r_0$ ($r_+$ and $r_-$) as a function of $\sqrt{s}$ are shown in Fig.~\ref{sin1 plot} and Fig. \ref{r plot}. We can see that the $\rho-\omega$ mixing mechanism produces a large $\sin\delta_0$ ($\sin\delta_+$ and $\sin\delta_-$) at the $\omega$ resonance. As can be seen from Fig.~\ref{sin1 plot}, the plots vary sharply in the cases of $\sin\delta_0$, $\sin\delta_+$ and $\sin\delta_-$. Meanwhile, $\sin\delta_0$ and $\sin\delta_-$ change weakly compared with the $\sin\delta_+$. It can be seen from Fig. \ref{r plot} that $r_{+}$ change more rapidly than $r_0$ and $r_-$ when the $\pi^+ \pi^-$ pairs in the vicinity of the $\omega$ resonance.

The Large Hadron Collider (LHC) is a proton-proton collider which has currently started at the European Organization for Nuclear Research (CERN). In order to achieve the required energy and luminosity, the technology and equipment has been upgraded many times. The LHC Run \uppercase\expandafter{\romannumeral1} first data-taking period lasted from 2010 to 2013 \cite{Aaij:2014jba}. In the next few years, there are two major detector (ATLAS and CMS) upgrades happening after
 Run \uppercase\expandafter{\romannumeral2} and Run \uppercase\expandafter{\romannumeral3}. With a series of upgrades and modifications, the LHC provides a TeV-level high energy frontier and an opportunity to further improve conformance testing of the CKM matrix. The production rates for heavy quark flavors will be much at the LHC, and the $b \bar b$ production cross section will be of the order of 0.5 mb, providing about $10^{12}$ bottom quark events per year \cite{Aaij:2014jba,Schopper:2006he}. The heavy flavour physics experiment is one of the main projects of LHC experiments. Especially, LHCb is a specialized B-physics experiment, designed primarily to precisly measure the parameters of new physics in $CP$ violation and rare decays in the interactions of beauty and
charm hadrons systems. Such studies can help to explain the Matter-Antimatter asymmetry of the Universe. Recently, the LHCb collaboration found clear evidence for direct $CP$ violation in some three-body decay channels in charmless decays of $B$ meson. Meanwhile, large $CP$ violation is obtained in $B^{\pm}\rightarrow \pi^{\pm}\pi^{+}\pi^{-}$ in the region $0.6$ GeV${^2}$$<m^{2}_{\pi^{+}\pi^{-}low}<0.8$ GeV${^2}$ and $m^{2}_{\pi^{+}\pi^{-}high}>14$ GeV${^2}$ \cite{Aaij:2014iva,dosReis:2016ayt}. A zoom of the $\pi^+ \pi^-$ invariant mass from the $B^{\pm}\rightarrow \pi^{\pm}\pi^{+}\pi^{-}$ decay process is shown  the region  $0.6$ GeV${^2}$$<m^{2}_{\pi^{+}\pi^{-}low}<0.8$ GeV${^2}$ zone in the Figure 8 of the Ref. \cite{dosReis:2016ayt}.  In addition, the branching fractions is probed in the $\pi^{+}\pi^{-}$ invariant mass
range $400 < m({\pi ^ + }{\pi ^ - }) < 1600$ MeV/$\rm {c^2}$ for $\bar{B}^0_{s}\rightarrow \pi^+\pi^-\phi$ \cite{Aaij:2016qnm}. In the next years, we expect the LHCb Collaboration to focus our prediction of $CP$ violation from the $\bar{B}^0_{s}\rightarrow \rho^0(\omega)\phi\rightarrow\pi^+\pi^-\phi$ decay process when the invariant
mass of $\pi^+\pi^-$ is in the vicinity of the $\rho$ resonance \cite{Aaij:2016qnm}. Theoretically, in order to achieve the current experiments on $b$-hadrons, which can only provide about $10^7$ $B \bar B$ pairs \cite{Guo:2008zzh}. Therefore,
it is very convenient to observe the $CP$ violation for $\bar{B}^0_{s}\rightarrow \rho^0(\omega)\phi\rightarrow\pi^+\pi^-\phi$ when the invariant masses of $\pi^{+}\pi^{-}$ pairs are in the vicinity
of the $\omega$ resonance at the LHC experiments.

\section{\label{sec:conclusion}Summary and conclusion}

In this paper, we study the $CP$ violation for the decay process of $\bar{B}^0_{s}\rightarrow \rho^0(\omega)\phi\rightarrow\pi^+\pi^-\phi$ due to the interference of $\rho$-$\omega$ mixing in perturbative QCD. It has been found that the $CP$ violation can be enhanced at the area of $\rho-\omega$ resonance. There is the resonance effect via $\rho$-$\omega$ mixing which can produce large strong phase in this decay process. As a result, one can find that the maximum $CP$ violation can reach $5.98\%$ when the invariant mass of the $\pi^+\pi^-$ pair is in the vicinity of the $\omega$ resonance.

In the calculation, we need the renormalization scheme independent Wilson coefficients for the tree and penguin operators at the scale $m_{b}$. The major uncertainties is from the input parameters. In particular,
these include the CKM matrix element parameters, the perturbative QCD approach and the hadronic parameters (the shape parameters, decay constants, the wave function and etc). We expect that our predictions will provide useful guidance for future investigations in $B_s$ decays.

\section*{Acknowledgements}
This work was supported by National Natural Science Foundation of China (Project Numbers 11605041), and the Research Foundation of the young core teacher from Henan province.

\appendix

\section*{\label{Appendix}  Appendix: Related functions defined in the text}

In this appendix we present explicit expressions of the factorizable and non-factorizable amplitudes in Perturbative QCD \cite{Ali:2007ff,Keum:2000ph,Keum:2000wi,Lu:2000em}. The factorizable amplitudes $F_{B_s\to \phi}^{LL,i}(a_i)$ (i=L,N,T) are written as

\begin{eqnarray}
  f_{M_2} F^{LL,L}_{B_s\to \phi} (a_i)&=&8\pi
  C_FM_{B_s}^4f_{M_2}\int^1_0dx_1dx_3\int^\infty_0b_1db_1b_3db_3
\phi_{B_s}(x_1,b_1)
  \Big\{a_i(t_a) E_e(t_a)
  \nonumber\\
  &&\times \Big[(1+x_3)\phi_3(x_3)+r_3(1-2x_3)(\phi_{3}^s(x_3)+\phi_3^t(x_3))
  \Big]h_e(x_1,x_3,b_1,b_3)
 \nonumber\\ && \;\;+2r_3\phi_3^s(x_3)a_i(t_a^\prime) E_e(t_a^\prime)h_e(x_3,x_1,b_3,b_1)
 \Big\},\label{ppefll}
\end{eqnarray}

\begin{eqnarray}
f_{M_2}F^{LL,N}_{B_s\to \phi}(a_i)&=&8\pi
C_FM_{B_s}^4f_{M_2}r_2\int^1_0dx_1dx_3\int^\infty_0b_1db_1b_3db_3\phi_{B_s}(x_1,b_1)
\Big\{h_e(x_1,x_3,b_1,b_3)\nonumber
\\
&&\times E_e(t_a)a_i(t_a)[\phi_{3}^T(x_3)+2r_3 \phi_{3}^v(x_3)+r_3 x_3
(\phi_{3}^v(x_3)-\phi_{3}^a(x_3))] \nonumber\\ &&
\;\;+r_3[\phi_{3}^v(x_3)+\phi_{3}^a(x_3)]E_e(t_a')a_i(t_a')h_e(x_3,x_1,b_3,b_1)
\Big\},
\end{eqnarray}

\begin{eqnarray}
f_{M_2}F^{LL,T}_{B_s \to {\phi}}( a_i)&=&16\pi
  C_FM_{B_s}^4f_{M_2}r_2\int^1_0dx_1dx_3\int^\infty_0b_1db_1b_3db_3\phi_{B_s}(x_1,b_1)
  \Big\{h_e(x_1,x_3,b_1,b_3)\nonumber\\
  &&\times [\phi_{3}^T(x_3)+2r_3 \phi_{3}^v(x_3)-r_3 x_3
  (\phi_{3}^v(x_3)-\phi_{3}^a(x_3))]E_e(t_a)a_i(t_a)
 \nonumber\\
 && \;\;+r_3[\phi_{3}^v(x_3)+\phi_{3}^a(x_3)]E_e(t_a')a_i(t_a')h_e(x_3,x_1,b_3,b_1)
 \Big\},
\end{eqnarray}
with the color factor ${C_F} = 3/4$, $f_{M_2}$, $f_{B_s}$ refer to the decay constant of $M_2$ ($\rho$ or $\omega)$ and $\bar{B}_{s}$ mesons and $a_i$ represents the corresponding Wilson coefficients for emission decay
 channels. In the above functions, $r_{2} (r_{3}) = m_{V_2}(m_{V_3}) /m_{B_s}$ and $\phi_{2} (\phi_{3}) = \phi_{\rho/\omega}(\phi_{\phi})$,
with $m_{B_s}$ and $m_{V_2}(m_{V_3})$ being the masses of the
initial and final states.

The non-factorizable amplitudes $M_{B_s \to {\phi}}^{LL,i}(a_i)$, and $M_{B_s \to {\phi}}^{SP,i}(a_i)$ (i=L,N,T) are written as
\begin{eqnarray}
 M_{B_s\to \phi}^{LL,L}(a_i)&=&32\pi
C_FM_{B_s}^4/\sqrt{6}\int^1_0dx_1dx_2dx_3\int^\infty_0b_1db_1b_2db_2
\phi_{B_s}(x_1,b_1)\phi_2(x_2)
\nonumber\\
&&\times
\Big\{\Big[(1-x_2)\phi_3(x_3)-r_3x_3(\phi_3^s(x_3)-\phi_3^t(x_3))\Big]
a_i(t_b)E_e^\prime(t_b)\nonumber\\
&&~\times h_n(x_1,1-x_2,x_3,b_1,b_2)+h_n(x_1,x_2,x_3,b_1,b_2)\nonumber\\
 &&\;\;\times\Big[-(x_2+x_3)\phi_3(x_3)+r_3x_3(\phi_3^s(x_3)+\phi_3^t(x_3))\Big]
 a_i(t_b^\prime) E_e^\prime(t_b^\prime)\Big\},\label{ppenll}
 \end{eqnarray}

\begin{eqnarray}
M_{B_s \to \phi}^{LL,N}(a_i)&=&32\pi
C_FM_{B_s}^4r_2/ \sqrt{6}\int^1_0dx_1dx_2dx_3\int^\infty_0b_1db_1b_2db_2\phi_{B_s}(x_1,b_1) \nonumber\\
 && \times
 \Big\{\left[x_2(\phi_2^v(x_2)+\phi_2^a(x_2))\phi_3^T(x_3)
 -2r_3(x_2+x_3)(\phi_2^v(x_2)\phi_3^v(x_3)+\phi_2^a(x_2)\phi_3^a(x_3))\right]
 \nonumber\\
  &&\;\;h_n(x_1,x_2,x_3,b_1,b_2)E_e'(t_b')a_i(t_b')\nonumber\\
  &&\;\;\;
+(1-x_2)(\phi_2^v(x_2)+\phi_2^a(x_2))\phi_3^T(x_3)
E_e'(t_b)a_i(t_b) h_n(x_1,1-x_2,x_3,b_1,b_2)\Big\},
\end{eqnarray}

\begin{eqnarray}
M_{B_s\to \phi}^{LL,T}(a_i)&=&64\pi C_FM_{B_s}^4r_2/
\sqrt{6}\int^1_0dx_1dx_2dx_3\int^\infty_0b_1db_1b_2db_2\phi_{B_s}(x_1,b_1)\Big\{
E_e'(t_b')a_i(t_b')\nonumber
\\
 &&\times \big[x_2(\phi_2^v(x_2)+\phi_2^a(x_2))\phi_3^T(x_3)
-2r_3(x_2+x_3)(\phi_2^v(x_2)\phi_3^a(x_3)\nonumber
\\
 && +\phi_2^a(x_2)\phi_3^v(x_3))\big] h_n(x_1,x_2,x_3,b_1,b_2)\nonumber
\\
 && +(1-x_2)[\phi_2^v(x_2)+\phi_2^a(x_2)]\phi_3^T(x_3)
E_e'(t_b)a_i(t_b) h_n(x_1,1-x_2,x_3,b_1,b_2)\Big\},
\end{eqnarray}

\begin{eqnarray} M^{SP,L}_{B_s\to \phi}( a_i) &=&32\pi C_F
M_{B_s}^4/\sqrt{6}\int^1_0dx_1dx_2dx_3\int^\infty_0b_1db_1b_2db_2
\phi_{B_s}(x_1,b_1)\phi_2(x_2)
\nonumber\\
&&\times\Big\{
\Big[(x_2-x_3-1)\phi_3(x_3)+r_3x_3(\phi_3^s(x_3)+\phi_3^t(x_3))\Big]\nonumber\\
&&\times
a_i(t_b)E_e^\prime(t_b)h_n(x_1,1-x_2,x_3,b_1,b_2)+a_i(t_b^\prime)
E^\prime_e(t_b^\prime)\nonumber\\
&&\times
 \Big[x_2\phi_3(x_3)+r_3x_3(\phi_3^t(x_3)-\phi_3^s(x_3))\Big]h_n(x_1,x_2,x_3,b_1,b_2)\Big\}.
 \label{ppensp}
\end{eqnarray}

\begin{eqnarray} M^{SP,N}_{B_s\to \phi}(a_i) &=&32 \pi C_F
M_{B_s}^4/\sqrt{6}\int^1_0dx_1dx_2dx_3\int^\infty_0b_1db_1b_2db_2
\phi_{B_s}(x_1,b_1)r_2\nonumber \\
&&\times\Big\{ x_2
(\phi_2^v(x_2)-\phi_2^a(x_2))\phi_3^T(x_3)E_e'(t_b')a_i(t_b')
h_n(x_1,x_2,x_3,b_1,b_2)\nonumber\\
&&+h_n(x_1,1-x_2,x_3,b_1,b_2)[(1-x_2)(\phi_2^v(x_2)-\phi_2^a(x_2))\phi_3^T(x_3)\nonumber \\
&&\;\;\;-2r_3(1-x_2+x_3)
(\phi_2^v(x_2)\phi_3^v(x_3)-\phi_2^a(x_2)\phi_3^a(x_3))]E_e'(t_b)a_i(t_b)
\Big\} ,
 \end{eqnarray}

\begin{eqnarray}
 M^{SP,T}_{B_s\to \phi}(a_i) &=&64\pi C_F
M_{B_s}^4/\sqrt{6}\int^1_0dx_1dx_2dx_3\int^\infty_0b_1db_1b_2db_2\phi_{B_s}(x_1,b_1)r_2 \nonumber \\
&&\times\Big\{ x_2
(\phi_2^v(x_2)-\phi_2^a(x_2))\phi_3^T(x_3)E_e'(t_b')a_i(t_b')
h_n(x_1,x_2,x_3,b_1,b_2)\nonumber
\\
&&+h_n(x_1,1-x_2,x_3,b_1,b_2)[(1-x_2)(\phi_2^v(x_2)-\phi_2^a(x_2))\phi_3^T(x_3)\nonumber \\
&&\;\;\;-2r_3(1-x_2+x_3)
(\phi_2^v(x_2)\phi_3^a(x_3)-\phi_2^a(x_2)\phi_3^v(x_3))]E_e'(t_b)a_i(t_b)
\Big\}.
\end{eqnarray}

The hard scale t are chosen as the maximum of the virtuality of the internal momentum transition in the hard amplitudes, including $1/b_i$:
\begin{eqnarray}
t_a&=&\mbox{max}\{{\sqrt
{x_3}M_{B_s},1/b_1,1/b_3}\},\\
t_a^\prime&=&\mbox{max}\{{\sqrt
{x_1}M_{B_s},1/b_1,1/b_3}\},\\
t_b&=&\mbox{max}\{\sqrt
{x_1x_3}M_{B_s},\sqrt{|1-x_1-x_2|x_3}M_{B_s},1/b_1,1/b_2\},\\
t_b^\prime&=&\mbox{max}\{\sqrt{x_1x_3}M_{B_s},\sqrt
{|x_1-x_2|x_3}M_{B_s},1/b_1,1/b_2\},\\
t_c&=&\mbox{max}\{\sqrt{1-x_3}M_{B_s},1/b_2,1/b_3\},\\
t_c^\prime
&=&\mbox{max}\{\sqrt {x_2}M_{B_s},1/b_2,1/b_3\},\\
t_d&=&\mbox{max}\{\sqrt {x_2(1-x_3)}M_{B_s},
\sqrt{1-(1-x_1-x_2)x_3}M_{B_s},1/b_1,1/b_2\},\\
t_d^\prime&=&\mbox{max}\{\sqrt{x_2(1-x_3)}M_{B_s},\sqrt{|x_1-x_2|(1-x_3)}M_{B_s},1/b_1,1/b_2\}.
\end{eqnarray}

The function h, coming from the Fourier transform of hard part H, are written as \cite{Li:2001ay},
\begin{eqnarray}
h_e(x_1,x_3,b_1,b_3)&=&\left[\theta(b_1-b_3)I_0(\sqrt
x_3M_{B_s}b_3)K_0(\sqrt
x_3 M_{B_s}b_1)\right.\nonumber\\
&& \left.+\theta(b_3-b_1)I_0(\sqrt x_3M_{B_s}b_1)K_0(\sqrt
x_3M_{B_s}b_3)\right]K_0(\sqrt {x_1x_3}M_{B_s}b_1)S_t(x_3),
\end{eqnarray}
\begin{eqnarray}
h_n(x_1,x_2,x_3,b_1,b_2)&=&\left[\theta(b_2-b_1)K_0(\sqrt
{x_1x_3}M_{B_s}b_2)I_0(\sqrt
{x_1x_3}M_{B_s}b_1)\right. \nonumber\\
&&\;\;\;\left. +\theta(b_1-b_2)K_0(\sqrt
{x_1x_3}M_{B_s}b_1)I_0(\sqrt{x_1x_3}M_{B_s}b_2)\right]\nonumber\\
&&\times
\left\{\begin{array}{ll}\frac{i\pi}{2}H_0^{(1)}(\sqrt{(x_2-x_1)x_3}
M_{B_s}b_2),& x_1-x_2<0\\
K_0(\sqrt{(x_1-x_2)x_3}M_{B_s}b_2),& x_1-x_2>0
\end{array}
\right. ,
\end{eqnarray}

\begin{eqnarray}
h_a(x_2,x_3,b_2,b_3)&=&(\frac{i\pi}{2})^2
S_t(x_3)\Big[\theta(b_2-b_3)H_0^{(1)}(\sqrt{x_3}M_{B_s}b_2)J_0(\sqrt
{x_3}M_{B_s}b_3)\nonumber\\
&&\;\;+\theta(b_3-b_2)H_0^{(1)}(\sqrt {x_3}M_{B_s}b_3)J_0(\sqrt
{x_3}M_{B_s}b_2)\Big]H_0^{(1)}(\sqrt{x_2x_3}M_{B_s}b_2),
\end{eqnarray}
\begin{eqnarray}
h_{na}(x_1,x_2,x_3,b_1,b_2)&=&\frac{i\pi}{2}\left[\theta(b_1-b_2)H^{(1)}_0(\sqrt
{x_2(1-x_3)}M_{B_s}b_1)J_0(\sqrt {x_2(1-x_3)}M_{B_s}b_2)\right. \nonumber\\
&&\;\;\left.
+\theta(b_2-b_1)H^{(1)}_0(\sqrt{x_2(1-x_3)}M_{B_s}b_2)J_0(\sqrt
{x_2(1-x_3)}M_{B_s}b_1)\right]\nonumber\\
&&\;\;\;\times K_0(\sqrt{1-(1-x_1-x_2)x_3}M_{B_s}b_1),
\end{eqnarray}
\begin{eqnarray}
h_{na}^\prime(x_1,x_2,x_3,b_1,b_2)&=&\frac{i\pi}{2}\left[\theta(b_1-b_2)H^{(1)}_0(\sqrt
{x_2(1-x_3)}M_{B_s}b_1)J_0(\sqrt{x_2(1-x_3)}M_{B_s}b_2)\right. \nonumber\\
&&\;\;\;\left. +\theta(b_2-b_1)H^{(1)}_0(\sqrt
{x_2(1-x_3)}M_{B_s}b_2)J_0(\sqrt{x_2(1-x_3)}M_{B_s}b_1)\right]\nonumber\\
&&\;\;\;\times
\left\{\begin{array}{ll}\frac{i\pi}{2}H^{(1)}_0(\sqrt{(x_2-x_1)(1-x_3)}M_{B_s}b_1),&
x_1-x_2<0\\
K_0(\sqrt {(x_1-x_2)(1-x_3)}M_{B_s}b_1),&
x_1-x_2>0\end{array}\right. ,
\end{eqnarray}
where $J_0$ and ${Y}_0$ are the Bessel function with $H_0^{(1)}(z) = \mathrm{J}_0(z) + i\, \mathrm{Y}_0(z)$.

The threshold re-sums factor $S_t$ follows the parameterized \cite{Kurimoto:2001zj}
\begin{eqnarray}
S_t(x)=\frac{2^{1+2c}\Gamma(3/2+c)}{\sqrt \pi \Gamma(1+c)}[x(1-x)]^c,
\end{eqnarray}
where the parameter $c=0.4$. In the non-factorizable contributions, $S_t(x)$ gives a very small numerical effect to the amplitude~\cite{Li:2002mi}.
Therefore, we drop $S_t(x)$ in $h_n$ and $h_{na}$.

The evolution factors $E^{(\prime)}_e$ and $E^{(\prime)}_a$ entering in the expressions for the matrix elements are given by
\begin{eqnarray}
E_e(t)&=&\alpha_s(t) \exp[-S_B(t)-S_3(t)], \ \ \ \ E'_e(t)=\alpha_s(t) \exp[-S_B(t)-S_2(t)-S_3(t)]|_{b_1=b_3},\\
E_a(t)&=&\alpha_s(t) \exp[-S_2(t)-S_3(t)],\ \ \ \ E'_a(t)=\alpha_s(t) \exp[-S_B(t)-S_2(t)-S_3(t)]|_{b_2=b_3},
\end{eqnarray}
in which the Sudakov exponents are defined as
\begin{eqnarray}
S_B(t)&=&s\left(x_1\frac{M_{B_s}}{\sqrt2},b_1\right)+\frac{5}{3}\int^t_{1/b_1}\frac{d\bar \mu}{\bar\mu}\gamma_q(\alpha_s(\bar \mu)),\\
S_2(t)&=&s\left(x_2\frac{M_{B_s}}{\sqrt2},b_2\right)+s\left((1-x_2)\frac{M_{B_s}}{\sqrt2},b_2\right)+2\int^t_{1/b_2}\frac{d\bar \mu}{\bar
\mu}\gamma_q(\alpha_s(\bar \mu)),
\end{eqnarray}
where $\gamma_q=-\alpha_s/\pi$ is the anomalous dimension of the quark. The explicit form for the  function $s(Q,b)$ is:
\begin{eqnarray}
s(Q,b)&=&\frac{A^{(1)}}{2\beta_{1}}\hat{q}\ln\left(\frac{\hat{q}}{\hat{b}}\right)-\frac{A^{(1)}}{2\beta_{1}}\left(\hat{q}-\hat{b}\right)+
\frac{A^{(2)}}{4\beta_{1}^{2}}\left(\frac{\hat{q}}{\hat{b}}-1\right)
-\left[\frac{A^{(2)}}{4\beta_{1}^{2}}-\frac{A^{(1)}}{4\beta_{1}}
\ln\left(\frac{e^{2\gamma_E-1}}{2}\right)\right]
\ln\left(\frac{\hat{q}}{\hat{b}}\right)
\nonumber \\
&&+\frac{A^{(1)}\beta_{2}}{4\beta_{1}^{3}}\hat{q}\left[
\frac{\ln(2\hat{q})+1}{\hat{q}}-\frac{\ln(2\hat{b})+1}{\hat{b}}\right]
+\frac{A^{(1)}\beta_{2}}{8\beta_{1}^{3}}\left[
\ln^{2}(2\hat{q})-\ln^{2}(2\hat{b})\right],
\end{eqnarray}
where the variables are defined by
\begin{eqnarray}
\hat q\equiv \mbox{ln}[Q/(\sqrt 2\Lambda)],~~~ \hat b\equiv
\mbox{ln}[1/(b\Lambda)], \end{eqnarray} and the coefficients
$A^{(i)}$ and $\beta_i$ are \begin{eqnarray}
\beta_1=\frac{33-2n_f}{12},~~\beta_2=\frac{153-19n_f}{24},\nonumber\\
A^{(1)}=\frac{4}{3},~~A^{(2)}=\frac{67}{9}
-\frac{\pi^2}{3}-\frac{10}{27}n_f+\frac{8}{3}\beta_1\mbox{ln}(\frac{1}{2}e^{\gamma_E}),
\end{eqnarray}
with $n_f$ is the number of the quark flavors and $\gamma_E$ is the
Euler constant. We will use the one-loop expression of the running coupling constant.

\end{spacing}
\end{document}